\documentclass[11pt,a4paper]{article}

\usepackage{amsmath}
\usepackage{amsthm}
\usepackage{graphicx}
\usepackage{mathtools}
\usepackage{amssymb}
\usepackage[english]{babel}
\usepackage[numbers,sort&compress]{natbib}
\usepackage[colorlinks=true, allcolors=blue]{hyperref}
\usepackage{thmtools}
\usepackage[capitalize]{cleveref}
\usepackage{tikz}
\usepackage{xspace}
\usepackage{standalone}
\usepackage{complexity}
\usepackage[margin=1in]{geometry}
\usepackage{bbm}
\usepackage{todonotes}
\usepackage{bm}
\newcommand{\TwoAgentFigure}{%
\begin{figure}[t]
\centering
\begin{tikzpicture}[
  x=1.15cm, y=1cm, font=\small,
  retained/.style={draw=blue!70!black, fill=blue!12},
  remainder/.style={draw=orange!70!black, fill=orange!20},
  edge to favourite/.style={draw=black!65, line width=0.55pt}
]
  \draw[retained]  (0,0) rectangle (6.2,0.54);
  \draw[remainder] (6.2,0) rectangle (10,0.54);
  \node at (3.1,0.27) {$P_1$};
  \node at (8.1,0.27) {$X$};

  \foreach \x/\marklabel in {0/0,10/1} {
    \draw (\x,0.54) -- ++(0,0.12) node[above] {$\marklabel$};
  }
  \draw[thick] (6.2,0) -- (6.2,0.66) node[above] {$m$};

\end{tikzpicture}
\caption{The two agent case. 
The anchor $m$ splits the cake in two pieces.
The division $[0, m]$ is trivially robust.}
\label{fig:two-agent-rightmost-mark}
\end{figure}
}

\newcommand{\ThreeAgentFigure}{%
\begin{figure}[t]
\centering
\begin{tikzpicture}[
  x=1.15cm, y=1cm, font=\small,
  retained/.style={draw=blue!70!black, fill=blue!12},
  remainder/.style={draw=orange!70!black, fill=orange!20},
  edge to favourite/.style={draw=black!65, line width=0.55pt}
]
  \draw[retained]  (0,0) rectangle (3.2,0.54);
  \draw[retained]  (3.2,0) rectangle (7.6,0.54);
  \draw[remainder] (7.6,0) rectangle (10,0.54);
  \node at (1.6,0.27) {$P_1$};
  \node at (5.4,0.27) {$P_2$};
  \node at (8.8,0.27) {$X$};

  \foreach \x/\marklabel in {0/0,1.8/c_1,5.9/c_3,7.6/m,10/1} {
    \draw (\x,0.54) -- ++(0,0.12) node[above] {$\marklabel$};
  }
  \draw[thick] (3.2,0) -- (3.2,0.66) node[above] {$c_2=\ell(m)$};

  \coordinate (leftpiece)  at (1.6,0);
  \coordinate (rightpiece) at (5.4,0);
  \coordinate (agent1) at (1.6,-0.78);
  \coordinate (agent2) at (3.2,-0.78);
  \coordinate (agent3) at (5.4,-0.78);
  \draw[edge to favourite] (agent1) -- (leftpiece)
                           (agent2) -- (leftpiece)
                           (agent2) -- (rightpiece)
                           (agent3) -- (rightpiece);
  \foreach \p in {leftpiece,rightpiece,agent1,agent2,agent3}
    \fill (\p) circle[radius=1.4pt];
  \foreach \i in {1,2,3}
    \node[below=2pt] at (agent\i) {Agent \i};
\end{tikzpicture}
\caption{The three agent case.
For each anchor $m$, there is a robust division of $[0,m]$.
For each agent $i$, $c_i$ is the point such that the agent is indifferent between $[0, c_i]$ and $[c_i, m]$.
The cut $\ell(m)$ is the median mark $\ell=c_2$.
For each agent, we draw a line between the agent and her favourite piece among $P_1, P_2$.}
\label{fig:three-agent-median-mark}
\end{figure}
}

\let\vec\mathbf

\DeclarePairedDelimiter\abs{\lvert}{\rvert}%
\DeclarePairedDelimiter\norm{\lVert}{\rVert}%

\DeclarePairedDelimiter\ceil{\lceil}{\rceil}

\makeatletter
\@ifundefined{newcounteralias}{}{%
    \renewcommand\thmt@autorefsetup{%
        \@xa\def\csname\thmt@envname autorefname\@xa
        \endcsname\@xa{\thmt@thmname}%
    }%
}
\makeatother

\newtheorem{lemma}{Lemma}[section]
\newtheorem{proposition}[lemma]{Proposition}
\newtheorem{claim}[lemma]{Claim}
\crefname{claim}{claim}{claims}
\Crefname{claim}{Claim}{Claims}

\theoremstyle{definition}
\newtheorem{definition}[lemma]{Definition}

\newcommand{\boldl}{\bm{\ell}}

\newcommand{\ellTuple}{(\ell_1, \ell_2, \ell_3, \ell_4)}
\newcommand{\RR}{\mathbb{R}}
\newcommand{\ZZ}{\mathbb{Z}}
\newcommand{\aloc}{\mathbf{A}}

\title{Envy-Free Cake-Cutting for Five Agents}

\author{
\begin{tabular}{cc}
& \\
\textbf{Aris Filos-Ratsikas} & \textbf{Alexandros Hollender}\\
\small{Imperial College London} & \small{Technical University of Munich}\\
\href{mailto:aris.filos-ratsikas@ed.ac.uk}{\small{\texttt{aris.filos-ratsikas@ed.ac.uk}}} & \href{mailto:alexandros.hollender@tum.de}{\small{\texttt{alexandros.hollender@tum.de}}}\\
& \\
\textbf{Mohamad Latifian} & \textbf{Giannis Tyrovolas}\\
\small{University of Edinburgh} & \small{University of Oxford}\\
\href{mailto:mohamad.latifian@ed.ac.uk}{\small{\texttt{mohamad.latifian@ed.ac.uk}}}  & \href{mailto:giannis.tyrovolas@cs.ox.ac.uk}{\small{\texttt{giannis.tyrovolas@cs.ox.ac.uk}}}\\
& \\
\end{tabular}
}

\date{}

\begin{document}

\maketitle

\begin{abstract}
We study the envy-free cake-cutting problem, in which a divisible resource must be divided among a set of agents with heterogeneous preferences in a way that ensures that no agent envies the piece obtained by another. Remarkably, under some mild continuity assumptions on the agent valuations, it is known that an envy-free division where each agent receives a \emph{connected} piece of the cake is always guaranteed to exist.

When the valuations are monotone, prior work has shown that a connected $\varepsilon$-envy-free division can be obtained using $\poly(\log(1/\varepsilon))$ queries for up to four agents, leaving the case of five or more agents open. In this work, we develop a new simpler approach that yields such an efficient algorithm for five agents. Our algorithm uses $\mathcal{O}(\log^3(1/\varepsilon))$ queries and also yields a novel proof of existence for five agents that does not rely on any fixed-point theorems beyond the intermediate value theorem.
\end{abstract}

\vfill

{\small
\paragraph{\bf AI usage disclosure.} The algorithm for five agents was obtained through interactions with ChatGPT 6 Astra on the 30th and 31st of October 2026. The authors then spent several weeks working on the presentation of the result. In particular, they simplified the approach to make it more intuitive and technically as simple as possible, while at the same time extracting the fundamental principles underpinning the algorithm, which they present separately as an existence proof. The entirety of the present paper was written by the authors, with AI used only for copy-editing and figures.

\section{Introduction}
Social choice studies how collective decisions should reflect individual preferences. Within this area, fair division addresses the allocation of resources among $n$ agents whose preferences may conflict~\citep{Procaccia16}. Even when a resource can be divided into equal physical shares, agents may disagree about which share is preferable: different locations in a plot of land or different periods of time need not be equally desirable. Fairness can be formalized in different ways. Proportionality, for instance, asks that each agent receive at least $1/n$ of her value for the whole resource. Another natural requirement is \emph{envy-freeness}: every agent should value her own share at least as much as anyone else's.

Cake cutting is a fundamental problem in fair division that captures this question for a divisible, heterogeneous resource~\citep{Steinhaus48}. The cake is modeled as the interval $[0,1]$, and a complete allocation assigns all of it to the agents. Shares may consist of several separated intervals or be required to be connected. In this work, we focus on the model where each agent receives a single interval. Such \emph{connected}, or \emph{contiguous}, shares are natural when dividing land or allocating uninterrupted access to a resource, and use only $n-1$ cuts for $n$ agents.

Classical results of \citet{Stromquist80}, \citet{Woodall80}, and \citet{Su99} establish the existence of contiguous envy-free allocations for any number of agents under standard continuity and hunger assumptions. Their proofs use fixed-point arguments or Sperner's lemma, but do not yield efficient algorithms. Since valuation functions can be arbitrarily complex, it is natural to elicit information through queries rather than assume a full description of each function. The Robertson--Webb model provides evaluation queries, which return an interval's value, and cut queries, which ask an agent to locate an endpoint attaining a specified value~\citep{WS07}.

Even for additive valuations, \citet{Stromquist08} proves that no finite Robertson--Webb protocol can always find an exact contiguous envy-free allocation for three or more agents. However, bounded exact protocols do exist when disconnected shares are allowed~\citep{AM16Any}. A standard way to bypass the connected impossibility is to allow a small amount of envy. An allocation $(A_1,\ldots,A_n)$ is \emph{$\varepsilon$-envy-free} if
\[
  v_i(A_i)\geq v_i(A_j)-\varepsilon
  \qquad\text{for all agents }i,j.
\]
The error is additive in each agent's own value, and exact envy-freeness requires zero error.

While more general valuation classes have also been studied in the literature, here we follow the monotone valuation model of \citet{HR25}. Valuations are monotone under inclusion, assign zero value to degenerate intervals, and are Lipschitz continuous in the endpoints: for a common bound $L\geq1$, moving the endpoints changes an interval's value by at most $L$ times their total displacement. Additivity is not required. Our algorithm uses only interval-value queries. As in \citet{HR25}, we state query bounds after normalizing the Lipschitz bound to one. For general $L$, this means replacing each valuation $v_i$ by $v_i/L$ and the target error $\varepsilon$ by $\varepsilon/L$.

Together with Lipschitz continuity, existence gives a simple baseline: a sufficiently fine grid contains an $\varepsilon$-envy-free allocation, which can be found by exhaustive search. For each fixed number of agents, this takes time polynomial in $1/\varepsilon$, but the degree grows with the number of agents. However, our goal is a query bound polynomial in $\log(1/\varepsilon)$, so that finer accuracy requires relatively few additional queries.

Such bounds are known for up to four agents. Binary search implements cut-and-choose with $O(\log(1/\varepsilon))$ value queries for two agents. \citet{DQS12} obtain $O(\log^2(1/\varepsilon))$ queries for three agents, and \citet{HR25} obtain $O(\log^3(1/\varepsilon))$ for four agents. These bounds apply to monotone Lipschitz interval valuations. Hollender and Rubinstein explicitly ask whether a polylogarithmic-query algorithm exists for five monotone agents, or whether a lower bound rules it out. We address this question.

\paragraph{Our contribution.}

We develop a common approach to contiguous approximate envy-free cake cutting for two through five agents. It recovers the known value-query bounds for two, three, and four agents and gives a deterministic algorithm using
$\poly(\log(1/\varepsilon))$ value queries for five agents under the assumptions above.

Most of the existing algorithms start with an initial state that is an equipartition of the cake for one particular agent.
Cut-and-choose relies on one agent's indifference between two pieces. Earlier three-agent procedures also exploit such indifferences~\citep{BB04}. In the four-agent construction of \citet{HR25}, one agent first divides the cake into four equally valued pieces. The cuts then move while she remains indifferent between either two or three of her favourite pieces and the other agents' preferences satisfy one of two specified patterns. A potential argument then shows that this continuous movement has to terminate at an envy-free division. Unfortunately, all attempts to use this approach for five agents fail. We instead take a completely different approach, which we detail in the technical overview below.

\paragraph{\bf Concurrent work.}
A very recent concurrent and independent work by \citet{mackenzie2026nearly} also provided an efficient algorithm for five agents. Their work includes additional results, most notably a lower bound for seven agents or more, showing that $\Omega(\poly(1/\varepsilon))$ queries are needed in the worst case. As mentioned in our AI disclosure statement, we obtained the results using ChatGPT~6 Astra on 30-31 October 2026. We then spent several weeks working on the write-up, and the preprint of \citet{mackenzie2026nearly} appeared during that time, on 17 November 2026. We have decided to upload the first version of our work without having looked at their manuscript beyond the abstract in order to ensure that it is truly independent. We aim to provide a comparison between the two algorithms in a future version.

\subsection{Technical Overview}

We begin our presentation with the three agent case to provide some intuition for our new ideas. We assume that the valuations satisfy the hungriness assumption.

The main idea behind our new approach is to introduce a main cut, which we will denote by $m$ and call the \emph{anchor} cut. For every possible way of fixing the anchor cut $m$ on the cake $[0,1]$, we will determine one additional cut and thus obtain a corresponding division of the cake into three pieces $P_1, P_2$ and $X$ with the following property:

\begin{quote}
\textbf{Robustness:} \textit{No matter which of the three agents departs, we can assign the pieces $P_1$, $P_2$ to the remaining two agents so that they do not envy each other.}
\end{quote}

We emphasize that the pieces $P_1$, $P_2$ have to be chosen \emph{before} some agent leaves. So it must be possible to assign the pieces no matter which agent departs. Furthermore, note that the robustness property does not say anything about how the agents value the piece $X$. Indeed, it only requires that the remaining two agents be envy-free among their allocated pieces $P_1$, $P_2$. Whether they value the piece $X$ more or less than their allocated piece is irrelevant.

So, assume for now that for each position of the anchor cut $m \in [0,1]$, we have the corresponding pieces $P_1(m)$, $P_2(m)$, and $X(m)$ satisfying the robustness property. Furthermore, assume that we have also managed to enforce the following two additional properties:

\begin{quote}
\textbf{Continuity in $\bm{m}$:} \textit{The endpoints of the pieces $P_1(m)$, $P_2(m)$ and $X(m)$ are continuous functions of $m$.}
\end{quote}

\begin{quote}
\textbf{Boundary conditions:} \textit{At $m=0$, all agents prefer piece $X(m)$ to any of the pieces $P_1(m)$, $P_2(m)$. At $m=1$, all agents prefer some piece among $P_1(m)$, $P_2(m)$ to piece $X(m)$.}
\end{quote}

Now, it is easy to use these properties to establish the existence of an envy-free division. Indeed, since the valuation functions are continuous and the pieces are continuous in $m$, we can apply the intermediate value theorem to obtain that there exists $m^* \in [0,1]$ such that:
\begin{itemize}
\item All agents weakly prefer some piece among $\{P_1(m^*),\allowbreak P_2(m^*)\}$ to piece $X$.
\item Some agent $i$ is indifferent between piece $X$ and its favourite piece among $\{P_1(m^*),\allowbreak P_2(m^*)\}$.
\end{itemize}
Note that we crucially used the boundary conditions at $m=0$ and $m=1$ here.

Finally, we argue that the division into pieces $P_1(m^*)$, $P_2(m^*)$ and $X(m^*)$ is envy-free. Indeed, first assign piece $X$ to agent $i$. This agent will not envy any other agent, since by the second bullet above, piece $X$ is one of its favourite pieces. Then, for the remaining agents, by the robustness property, we can assign the pieces $P_1(m^*)$, $P_2(m^*)$ to them so that they do not envy each other. Furthermore, they also do not envy agent $i$, since by the first bullet above, they weakly prefer their piece to piece $X$. Thus, we have obtained an envy-free division.

In order to complete this argument, it remains to show that for any anchor cut $m \in [0,1]$ there indeed exist pieces $P_1$, $P_2$, and $X$ that satisfy the robustness property, as well as continuity in $m$ and the boundary conditions. Next, we will provide some intuition for why this is the case, starting with the simpler setting where we only have two agents.

Before we proceed, let us also briefly discuss the algorithmic implications of what we have seen so far. The fact that we have only used the intermediate value theorem means that we can hope to use this approach to obtain an efficient algorithm. Indeed, we can simply use binary search on $m$ to identify a value that is sufficiently close to an exact solution anchor $m^*$. For this to work, we need to be able to efficiently compute, given $m$, the pieces $P_1(m)$, $P_2(m)$, and $X$, or to be able to approximate them sufficiently well.

\paragraph{\bf Two agents.}
In this case, given $m \in [0,1]$, we simply let $P_1(m) := [0,m]$ and $X(m) := [m,1]$. Robustness holds trivially, since if one of the two agents leaves, the other agent will be happy with piece $P_1$, as there will be no other agent left that could be envied. Continuity in $m$ of the pieces is also immediate from their construction. Finally, it is easy to see that the boundary conditions are also satisfied. This is illustrated in \Cref{fig:two-agent-rightmost-mark}.

\TwoAgentFigure

As a result, the intermediate value theorem guarantees that there exists $m^*$ such that both agents weakly prefer piece $[0,m^*]$ to piece $[m^*,1]$, and at least one of them is indifferent between the two. We can thus assign the pieces in an envy-free way. This approach is similar, but not identical, to cut-and-choose. In cut-and-choose we pick an agent and ask them to cut the cake into two pieces of equal value, then letting the other agent choose among those two pieces first. Here, we can instead think of the protocol as asking both agents to mark their halfway point on the cake, and then picking the rightmost such mark as the cut $m^*$. In any case, this yields an algorithm that finds an $\varepsilon$-envy-free allocation using $\mathcal{O}(\log(1/\varepsilon))$ queries.

\paragraph{\bf Three agents.}
Given $m \in [0,1]$, we let $P_1(m) := [0,\ell]$, $P_2(m) := [\ell,m]$, and $X(m) := [m,1]$, where $\ell$ is set such that at least two agents weakly prefer piece $P_1$ to $P_2$, and at least two agents weakly prefer piece $P_2$ to $P_1$. Since we have assumed that the valuations are hungry, $\ell$ is in fact uniquely determined by this condition. The condition also immediately implies robustness, i.e., no matter which agent leaves, we can assign $P_1$ and $P_2$ to the remaining two agents so that they do not envy each other. This is illustrated in \Cref{fig:three-agent-median-mark}.

\ThreeAgentFigure

Another equivalent way of describing the choice of $\ell$ is as follows. For each agent $i \in \{1,2,3\}$, let $c_i$ be the unique cut such that agent $i$ is indifferent between pieces $[0,c_i]$ and $[c_i,m]$. Then, set $\ell$ to be the median of $c_1$, $c_2$, and $c_3$. From this we also obtain that $\ell$ is continuous as a function of $m$, since it is not hard to see that the $c_i$ are continuous functions of $m$. As a result, the pieces are all continuous in $m$, as desired. Finally, the boundary conditions at $m=0$ and $m=1$ are trivially satisfied.

We thus obtain an existence proof of an envy-free division that uses the intermediate value theorem twice: once to obtain $\ell$ given any fixed $m$, and once to obtain the solution anchor cut $m^*$. This naturally yields a nested binary search algorithm. An outer binary search operates on the value of $m$ and uses an inner binary search to identify (or rather approximate) the value of $\ell$ corresponding to the current $m$. Thus, this gives rise to an algorithm for finding an $\varepsilon$-envy-free allocation using $\mathcal{O}(\log^2(1/\varepsilon))$ queries.

\paragraph{\bf Four agents.}
Now, given $m \in [0,1]$, we set $P_1(m) := [0,\ell]$, $P_2(m) := [\ell,m]$, $P_3(m) := [m,r]$, and $X(m) := [r,1]$. Here, $\ell$ and $r$ are chosen such that the robustness condition holds, i.e., no matter which agents leaves, we can assign the pieces $P_1, P_2, P_3$ to the remaining three agents so that they do not envy each other.

Unlike in the case of fewer agents, for four agents it is no longer true that such $\ell$ and $r$ necessarily exist for every $m \in [0,1]$. This is not too hard to see. For example, when $m$ is close to $1$, the piece $P_3$ is very small and it is impossible to ensure that some agent will weakly prefer it over the best piece among $P_1$ and $P_2$. Intuitively, for such cases, namely when no choice of $P_3$ allows us to enforce the robustness property, we would like to be able to fix $P_3 := [m,1]$ and to instead allow a new unused interval to appear between the pieces $P_1$ and $P_2$. Thus, the pieces $P_1$ and $P_2$ could then both become small despite the interval $[0,m]$ being large. As it turns out, the solution to this issue also allows us to resolve the five agent setting, so we present it next directly for five agents.

\paragraph{\bf Five agents.}
In order to resolve the issue, we allow a second unused piece. Namely, given $m \in [0,1]$, we consider a division of the cake parameterized by $\ell_1, \ell_2, \ell_3, \ell_4$ with the main pieces $P_1(m) := [0,\ell_1]$, $P_2(m) := [m-\ell_2,m]$, $P_3(m) := [m,m+\ell_3]$, $P_4(m) := [1-\ell_4,1]$ and the additional pieces $L(m) := [\ell_1,m-\ell_2]$ and $R(m) := [m+\ell_3,1-\ell_4]$. We also enforce feasibility, namely
\[\ell_1 + \ell_2 \leq m \qquad \text{and} \qquad \ell_3 + \ell_4 \leq 1-m\]
as well as maximality, which additionally requires that one of these two inequalities holds with equality. Note that this means that at least one of $L(m)$ or $R(m)$ has to be empty at any given anchor cut $m$. This will be crucial for our argument. See \cref{fig:geometry} for an illustration.

With this updated definition, we recover existence and uniqueness. Namely, for any given $m \in [0,1]$, there exist unique $\ell_1, \ell_2, \ell_3, \ell_4$ that are maximal feasible such that the robustness property is satisfied for the pieces $P_1, P_2, P_3, P_4$. This is non-trivial and we will provide some intuition about why it holds below. When establishing the existence and uniqueness of the $\ell_1, \ell_2, \ell_3, \ell_4$, we also show that they are continuous functions of $m$, which is also used in a crucial way.

Let us now see how we can use this to prove existence of an envy-free division. The idea is to treat the right piece $R$ in the same way as we treated $X$ before, and to essentially ignore the left piece $L$. At $m=0$, we necessarily have that $R = [0,1]$ and all the other pieces are empty. Thus, all agents prefer $R$ to any of the pieces $P_1, P_2, P_3, P_4$. At $m=1$, the piece $R$ is necessarily empty and so all agents prefer some other piece to it (namely, piece $L$ which is $[0,1]$). As a result, by the intermediate value theorem, there exists a position of the anchor cut $m^*$ such that all agents weakly prefer some other piece to piece $R$, and some agent $i$ is indifferent between its favourite piece and piece $R$.

The crucial observation is that since $R$ is among the favourite pieces of agent $i$, it cannot be empty (by hungriness of the valuations). But this means that piece $L$ must necessarily be empty. As a result, we can assign piece $R$ to agent $i$, and by the robustness property we can assign the pieces $P_1, P_2, P_3, P_4$ to the remaining four agents so that they do not envy each other. This yields an envy-free division of the cake.

\paragraph{\bf Existence of $\bm{\ell_1, \ell_2, \ell_3, \ell_4}$.}
Here we give some intuition to explain why each $m \in [0,1]$ there exist unique maximal feasible $\ell_1, \ell_2, \ell_3, \ell_4$ so that the robustness property holds. Assume that we have fixed $\ell_1$ to some very small value. Then, we can show that there exist unique values of $\ell_2, \ell_3$ and $\ell_4$ such that the robustness property is satisfied. These values of $\ell_2, \ell_3$ and $\ell_4$ will be very small, and so the maximality condition will not hold. Now imagine that we increase $\ell_1$, while updating $\ell_2, \ell_3$ and $\ell_4$ to maintain robustness. Then, $\ell_2, \ell_3$ and $\ell_4$ will also increase along with $\ell_1$. Thus, at some point one of the two inequalities $\ell_1 + \ell_2 \leq m$ or $\ell_3 + \ell_4 \leq 1-m$ will become tight. At that point we will also have maximality (i.e., at least one of $L$ or $R$ is empty), and thus the desired division.

In \cref{sec:existence} we provide an argument using only the intermediate value theorem showing that given $\ell_1$, there exist unique values of $\ell_2, \ell_3$ and $\ell_4$ such that the robustness property holds. This in turn, provides an algorithm using $\mathcal{O}(\log^5(1/\varepsilon))$ queries due to three nested binary searches to determine $\ell_2, \ell_3, \ell_4$, one more for $\ell_1$, and the outer binary search over $m$. Instead of this, we show that $\mathcal{O}(\log^2(1/\varepsilon))$ queries suffice to find $\ell_1, \ell_2, \ell_3, \ell_4$ for a given $m$, using an enumeration over the constant number of possible indifference graphs. This gives the final bound for the algorithm of $\mathcal{O}(\log^3(1/\varepsilon))$ queries.

\subsection{Related work}

\paragraph{Exact protocols and disconnected shares.}
For additive valuations, \citet{BT95} and \citet{RW97} give finite exact envy-free protocols allowing disconnected shares, but their query counts are not bounded solely by the number of agents. \citet{AM16} obtain a bound of $584$ queries for four agents. Their protocol for arbitrary $n$~\citep{AM16Any} uses at most $n^{n^{n^{n^{n^n}}}}$ queries, a tower of six $n$'s. Recent deterministic protocols use $2^{O(n\log n)}$ queries with an elementary proof~\citep{Psomas26} and $n^{O(1)}2^n$ queries~\citep{YB26}. \citet{MS26} give a randomized protocol using $\widetilde O(n^{30})$ queries in expectation for every valuation profile. Moving-knife procedures offer another route to exact division~\citep{BTZ97,BB04,SW09}. In particular, the five-person procedure of \citet{SW09} has a polylogarithmic approximate discretization for additive valuations with bounded density, but permits disconnected shares.

\paragraph{Approximation and hardness.}
For a variable number of agents, \citet{GHS20} study connected approximation and constrained variants, and \citet{BK23} obtain additive envy at most $1/4+o(1)$ for normalized additive valuations, together with a multiplicative guarantee. Our focus is arbitrarily small additive error for a fixed population. Without monotonicity, query lower bounds polynomial in $1/\varepsilon$ hold already for four agents~\citep{HR25} and three agents~\citep{HMR26}. These lower bounds do not apply to our valuation class. For normalized additive valuations given by rational piecewise-constant densities, \citet{KRS26} prove PPAD-completeness of both exact and approximate complete connected envy-freeness. Their reduction uses a growing number of agents. It therefore does not establish hardness for five or any other fixed number of monotone agents. Beyond five agents, the cited results leave the polylogarithmic-query question open for every fixed number of monotone agents.

\paragraph{Query access and communication.}
Our monotone valuation model does not require additivity, but assumes a known Lipschitz bound and uses only value queries. In the standard Robertson--Webb model for additive valuations, cut queries are also available and no density bound is imposed. In that model, \citet{BN22} give a tight deterministic $\Theta(\log(1/\varepsilon))$ bound for three agents and an $O(n/\varepsilon)$ query upper bound for connected approximate envy-freeness for arbitrary $n$. These results differ from ours in both the valuation class and the permitted queries. The general query bound counts preference elicitation, while the subsequent search need not take polynomial time. It is therefore compatible with the variable-agent PPAD-hardness result. Communication complexity instead measures the bits exchanged by agents who know their valuations~\citep{BN19}. A bound on value queries, whose answers may be arbitrary real numbers, does not by itself give the same bound on communicated bits.

\paragraph{Leaving agents and other requirements.}
In the leaving-agent problem, the cake is cut into $n-1$ pieces before it is known which of the $n$ agents will depart. Whichever agent leaves, the remaining agents must admit an envy-free assignment of those pieces. \citet{MS19} prove existence under standard hunger and closedness assumptions. This is the robustness property used for our retained intervals. Secretive-agent results instead prepare $n$ pieces without consulting one agent. After she chooses a favourite, the other agents can receive pieces they weakly prefer to every other piece~\citep{AFP18}. \citet{PZ22} study both settings, including extensions allowing preferences for degenerate pieces under additional assumptions and restrictions on the possible assignments. These existence results do not supply the stability and geometric search needed by our algorithm. Other work studies exact contiguous envy-freeness with free disposal~\citep{SHA16} and truthful mechanisms for restricted valuations~\citep{AFGST17}.
 
\section{Preliminaries}

For positive integers $n$, we will abbreviate the set $\{1, \ldots, n\}$ as $[n]$.
We will write $\norm{\cdot}_\infty$ for the supremum norm. That is, for a vector $\vec{x} \in \RR^n$, $\norm{\vec{x}}_\infty = \max_{j \in [n]} \abs{x_j}$.

The cake is modelled as the interval $[0,1]$. There are $n$ agents, each with a valuation function that assigns a value to every interval of the cake. Namely, for each $i \in [n] := \{1,2, \dots, n\}$, there is a function $v_i \colon [0,1]^2 \rightarrow [0, 1]$. The value of interval $[\ell,r]$ for agent $i$ is then given by $v_i(\ell,r)$. We require that if $\ell \geq r$, then $v_i(\ell, r) = 0$. Furthermore, we assume that the valuation functions also satisfy the following two properties:

\begin{itemize}
    \item For some constant $L > 0$, the $v_i$ are \emph{$L$-Lipschitz-continuous}, i.e., for all $(\ell, r), (\ell', r')\in [0,1]^2$, 
    $$\abs{v_i(\ell, r) - v_i(\ell', r')} \leq L(\abs{\ell - \ell'} + \abs{r - r'}).$$
    \item The $v_i$ are \emph{monotone}. That is, for all $[\ell, r] \subseteq [\ell', r']$, $v_i(\ell, r) \leq v_i(\ell', r')$.
\end{itemize}

We will refer to a contiguous division of the cake as a collection of $k$ intervals $P_1, \ldots, P_k$ with disjoint interiors.
If the $P_j$ partition the cake, i.e., $\cup_{j \in [k]} P_j = [0,1]$, then we refer to this division as a complete division. A (partial) allocation of pieces to a set of agents $S$ is an injective mapping from agents in $S$ to pieces. We will write $A_i$ for the piece that agent $i$ receives.

\begin{definition}
    Let $S$ be a subset of agents. An allocation $\aloc = \{A_i\}_{i \in S}$ is \emph{envy-free} among agents in $S$, if for all $i, j \in S$, $v_i(A_i) \geq v_i(A_j)$. It is \emph{$\varepsilon$-envy-free} if the inequality holds up to an additive $\varepsilon$ term: for all $i, j \in S$, $v_i(A_i) +\varepsilon\geq v_i(A_j)$.
\end{definition}

Notice here that we allow $\aloc$ to not partition $[0,1]$.

\paragraph{\textbf{Normalization.}}
Without loss of generality, we can assume that the valuations are $1$-Lipschitz-continuous, i.e., $L = 1$. Indeed, this can be enforced by replacing $v_i$ by $v_i/L$, and the approximate envy-freeness target $\varepsilon$ by $\varepsilon/L$. This will allow us to express computational bounds only in terms of $\varepsilon$.

\paragraph{\textbf{Query complexity.}}
We consider algorithms that can access the valuation functions by querying their value.\footnote{These are often called \emph{value queries} in the relevant literature.} A query consists of two endpoints $\ell, r$ and the identity of an agent $i \in [n]$, and the response is the value of the interval $[\ell,r]$ for that agent, i.e., $v_i(\ell,r)$. The performance of an algorithm is measured in terms of the number of queries needed to output an $\varepsilon$-envy-free allocation. It is not hard to see that this problem can be solved using a brute-force algorithm that uses $\poly(1/\varepsilon)$ queries. We consider an algorithm to be efficient if it uses $\poly(\log(1/\varepsilon))$ queries.

Although the model only requires an algorithm to be query-efficient, but not necessarily time-efficient, the algorithm we present in this paper also runs in polynomial time in the corresponding white-box model. See also \cite{HollenderR23-cake-four} for more details and a comparison to other models.

\paragraph{\textbf{Hungriness.}}
In the cake-cutting problem, it can be convenient to assume that the valuations are not only monotone, but strictly monotone. This is usually called the \emph{hungriness} condition. We say that $v_i$ is \emph{hungry}, if for all $[\ell, r] \subsetneqq [\ell', r']$, we have $v_i(\ell, r) < v_i(\ell', r')$. For computational purposes, the following stronger version of this condition is very useful. We say that $v_i$ is \emph{$\gamma$-strongly hungry} for some $\gamma > 0$, if for all $[\ell, r] \subseteq [\ell', r']$
$$v_i(\ell', r') \geq \gamma(\ell - \ell') + \gamma(r' - r) + v_i(\ell, r).$$

We do not need to assume that the valuations are hungry or strongly hungry. Instead, it is possible to ensure that this is the case by a simple preprocessing step. This step slightly alters the valuation functions, but as we are only interested in finding approximately envy-free allocations, and not exact ones, this is not an issue. In more detail, given $\gamma \in (0,1)$, we can simply replace each valuation function $v_i$, by $v_i'$, where $v_i'(\ell,r) := (1-\gamma)v_i(\ell,r) + \gamma |r-\ell|$. It is easy to see that $v_i'$ is $\gamma$-strongly hungry, and that it remains $1$-Lipschitz-continuous like $v_i$. Furthermore, any $\delta$-envy-free division with respect to the $v_i'$ yields a $(\delta + 2\gamma)$-envy-free allocation with respect to the original valuations $v_i$. Thus, by picking $\gamma$ and $\delta$ appropriately, we can ensure that we obtain the desired $\varepsilon$-envy-free allocation at the end.

\section{Existence Proof for Five Agents}
\label{sec:existence}

We start by presenting a novel constructive existence proof for an exact envy-free division for five agents.
\begin{restatable}{theorem}{theoremExistence}
    \label{theorem:existence-of-ef-5}
    Consider an instance of cake-cutting with five agents with valuations $v_i$ that are monotone and Lipschitz continuous. 
    Then, there exists an envy-free allocation with contiguous pieces that completely allocates the cake to five agents.
\end{restatable}

From the proof, we will be able to extract an algorithm that efficiently computes an approximately envy-free division.
We start by setting up some useful definitions.

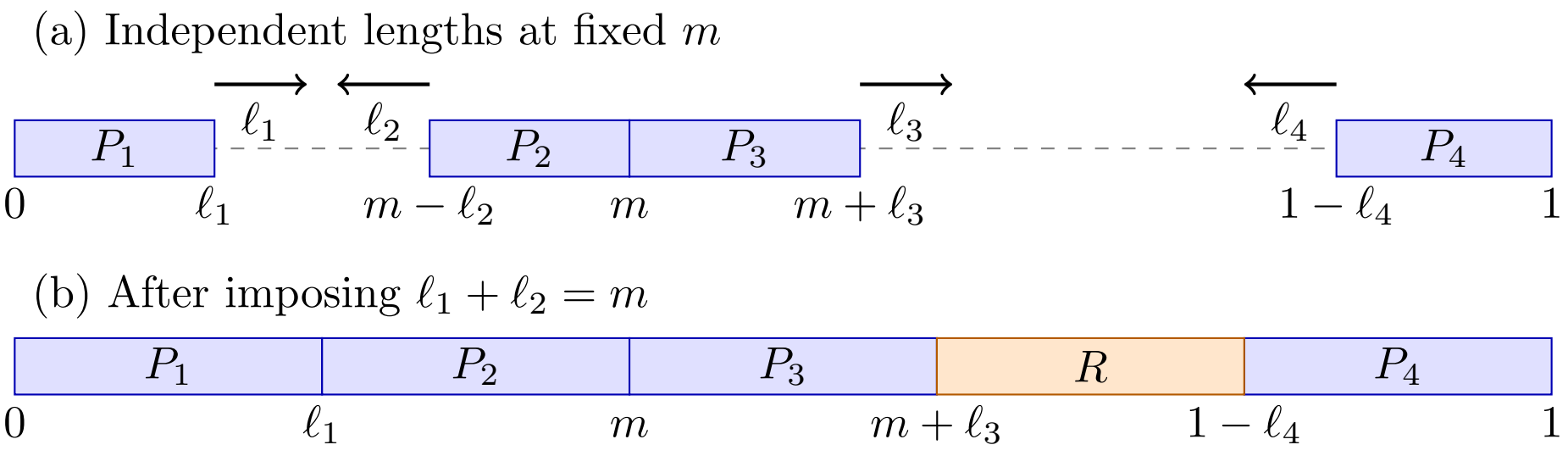
\begin{figure}[t]
\centering
\begin{tikzpicture}[
  x=1.2cm, y=1cm, font=\small,
  retained/.style={draw=blue!70!black, fill=blue!12},
  remainder/.style={draw=orange!70!black, fill=orange!20}
]
  \node[anchor=west] at (0,1.5)
    {(a) Independent lengths at fixed $m$};
  \draw[gray,dashed] (0,0.6) -- (10,0.6);

  \foreach \a/\b/\j in {0/1.3/1,2.7/4/2,4/5.5/3,8.6/10/4} {
    \draw[retained] (\a,0.38) rectangle (\b,0.82);
    \node at ({(\a+\b)/2},0.6) {$P_{\j}$};
  }

  \draw[->,thick] (1.3,1.10) -- node[midway, below] {$\ell_1$}  (1.9,1.10);
  \draw[->,thick] (2.7,1.10) -- node[midway, below] {$\ell_2$} (2.1,1.10);
  \draw[->,thick] (5.5,1.10) -- node[midway, below] {$\ell_3$} (6.1,1.10);
  \draw[->,thick] (8.6,1.10) -- node[midway, below] {$\ell_4$} (8.0,1.10);

  \foreach \x/\lab in {
    0/0,1.3/\ell_1,2.7/m-\ell_2,4/m,
    5.5/m+\ell_3,8.6/1-\ell_4,10/1
  } {
    \node[anchor=base] at (\x,0.06) {$\lab$};
  }

  \node[anchor=west] at (0,-0.55)
    {(b) After imposing $\ell_1+\ell_2=m$};

  \foreach \a/\b/\j in {0/2/1,2/4/2,4/6/3,8/10/4} {
    \draw[retained] (\a,-1.32) rectangle (\b,-0.88);
    \node at ({(\a+\b)/2},-1.1) {$P_{\j}$};
  }

  \draw[remainder] (6,-1.32) rectangle (8,-0.88);
  \node at (7,-1.1) {$R$};

  \foreach \x/\lab in {
    0/0,2/\ell_1,4/m,6/m+\ell_3,8/1-\ell_4,10/1
  } {
    \node[anchor=base] at (\x,-1.65) {$\lab$};
  }
\end{tikzpicture}
\caption{The four retained pieces and their growth directions.
Closing the left gap leaves the interval
$R=[m+\ell_3,1-\ell_4]$, which is assigned to the fifth agent at contact.}
\label{fig:geometry}
\end{figure}

\begin{definition}\label{def:feasible}
For any anchor $m \in [0,1]$, we say that the tuple $(\ell_1, \ell_2, \ell_3, \ell_4) \in [0,1]^4$ is \emph{feasible}, if
\[\ell_1 + \ell_2 \leq m \qquad \text{and} \qquad \ell_3 + \ell_4 \leq 1-m.\]
A feasible tuple is \emph{maximal} if one of the inequalities is tight.
\end{definition}

Given an anchor $m \in [0,1]$ and a tuple $\boldl = (\ell_1, \ell_2, \ell_3, \ell_4)$, we construct four pieces.
We use the following shorthands: $P_1 = [0, \ell_1], P_2 = [m - \ell_2, m], P_3 = [m, m + \ell_3], P_4 = [1 - \ell_4, 1]$.
Under this interpretation, a tuple $\boldl$ is feasible if pieces $P_j$ are subsets of $[0,1]$ and their interiors are disjoint. A feasible tuple is maximal if pieces $P_j$ cover all of the cake except for at most one non-empty interval.
Also, as a matter of notation, a tuple $\boldl^*$ similarly defines pieces $P_1^* = [0, \ell^*_1], \ldots, P_4^*= [1-\ell^*_4, 1]$.

\begin{definition}\label{def:robust}
For any $m \in [0,1]$, we say that a feasible tuple $(\ell_1, \ell_2, \ell_3, \ell_4)$ is \emph{robust}, if, no matter which agent departs, we can assign the pieces $P_1, P_2, P_3, P_4$ to the remaining four agents so that they do not envy each other. More generally, we say that the tuple is \emph{$\delta$-robust} for some $\delta \geq 0$, if, no matter which agent departs, we can assign the pieces $P_1, P_2, P_3, P_4$ to the remaining four agents so that they do not envy each other by more than $\delta$.
\end{definition}

\begin{definition}
    For $m \in [0, 1]$, $\delta \geq 0$ and a feasible tuple $\boldl$, we define the $\delta$-favourite graph $G_\delta(m, \boldl)$.
    This is a bipartite graph with vertices $[5] \sqcup \{P_1, \ldots, P_4\}$.
    There is an edge from vertex $i$ to vertex $P_j$ if and only if $v_i(P_j) + \delta \geq \max_{k \in [4]}v_i(P_k)$.
    If $m$ is clear from context, we will use the shorthand $G_\delta(\boldl)$.
\end{definition}

If $v_i(P_j) + \delta \geq \max_{k \in [4]}v_i(P_k)$ we will say that piece $P_j$ is a $\delta$-favourite piece for agent $i$.
Let $S$ be a subset of the vertex set on the same side of the bipartite graph. Let $N(S)$ denote its set of neighbours. We can now describe $\delta$-robustness in the language of graphs.
\begin{lemma}
    \label{lemma:robustness-is-equivalent-to-matching}
    Let $\boldl$ be a feasible tuple.
    Then, $\boldl$ is $\delta$-robust if and only if for all $i \in [5]$ and for all $S \subseteq [5] \setminus \{i\}$, $\abs{N(S)} \geq \abs{S}$ in the graph $G_\delta(\boldl)$.
\end{lemma}

\begin{proof}
    This is a direct application of Hall's condition. Any $\delta$-envy-free allocation is a matching of pieces to agents for which the pieces are $\delta$-favourite. Once an agent $i$ departs, such a matching exists if and only if for all $S \subseteq [5] \setminus\{i\}$, $\abs{S} \leq \abs{N(S)}$.
\end{proof}

\begin{lemma}
    If $\boldl$ is a $\delta$-robust division, then $G_\delta(\boldl)$ is connected.
\end{lemma}
\begin{proof}
    Suppose for a contradiction that $G_\delta(\boldl)$ is disconnected.
    Let $S \subseteq [5]$ be the set of agents in a maximal connected component.
    Then, if $\abs{S} \leq 4$ there is $i \in [5] \setminus S$. From \Cref{lemma:robustness-is-equivalent-to-matching}, $\abs{N(S)} \geq \abs{S}$.
    Similarly, for $[5] \setminus S$, $\abs{N([5] \setminus S)} \geq \abs{[5]\setminus S}$.
    As the graph is disconnected and $S$ is maximal, $N([5] \setminus S)$ and $N(S)$ are disjoint. So, $\abs{N(S)} + \abs{N([5] \setminus S)} \leq 4$.
    Hence $\abs{S} + \abs{[5] \setminus S} = 5\leq 4$, a contradiction.
    So, $\abs{S} = 5$. 
    For $G_\delta(\boldl)$ to be $\delta$-robust, $N([5]) = \{P_1, P_2, P_3, P_4\}$ by \Cref{lemma:robustness-is-equivalent-to-matching}.
    Hence, the graph $G_\delta(\boldl)$ is connected.
\end{proof}

To prove the existence of a $\delta$-robust allocation we introduce a ``weight'' function.
Each agent $i$ can contribute $w_{ij}$ to the weight of piece $P_j$.
If a piece $P_j$ is the unique favourite of agent $i$, the weight $w_{ij}$ will be close to $1$.
More generally, if piece $P_j$ is one of $k$ joint favourite pieces, the weight $w_{ij}$ is close to $1/k$.
The weight of piece $P_j$, denoted by $W_j$ is the sum of the weights of the agents for that piece, i.e., $W_j = \sum_{i \in [5]} w_{ij}$.
In addition, we choose a weight function that is continuous.

\begin{definition}
    For a fixed $m$ and $\delta > 0$, consider a tuple $\boldl$. 
    Let $\exp(x) = 2^x$. For a piece $P_j$ and an agent $i$ let
    \begin{equation*}
        w_{ij}(m, \boldl) = \frac{\exp(\frac{5}{\delta} \cdot v_i(P_j))}{\sum_{k \in [4]}\exp(\frac{5}{\delta} \cdot v_i(P_k))}.
    \end{equation*}
    Then, let $W_j(m, \boldl) = \sum_{i \in [5]} w_{ij}(m, \boldl)$. When clear from context, we will suppress $m$ and write $W_j(\boldl)$ instead.
\end{definition}

\begin{lemma}
    For each agent $i$, $\sum_{j \in [4]} w_{ij}(\boldl) = 1$ and we have that $\sum_{j \in [4]} W_j(\boldl) = 5$.
\end{lemma}

\begin{proof}
    This can be directly observed from the definition of $w_{ij}$:

    \begin{align*}
        \sum_{j \in [4]}w_{ij}(m, \boldl) &= \frac{\sum_{j \in [4]}\exp(\frac{5}{\delta} \cdot v_i(P_j))}{\sum_{k \in [4]}\exp(\frac{5}{\delta} \cdot v_i(P_k))}\\
        &= 1.
    \end{align*}
    As there are 5 agents, we can deduce $\sum_{j \in [4]} W_{j}(\boldl) = 5$.
\end{proof}
\begin{lemma}
    \label{lemma:strict-monotonicity}
    Assume that the valuations $v_i$ are hungry. Then, the weight function $W_j(m, \boldl)$ is strictly increasing in $\ell_j$, strictly decreasing in $\ell_k$ for $k \neq j$ and continuous.
\end{lemma}

\begin{proof}
    Strict monotonicity follows directly from the fact that $W_j$ is the composition of strictly monotone functions. To prove this fact, we rely on the fact that $v_i$ are strictly monotonic.
    Similarly, as the denominator of $w_{ij}$ is always at least $4$, and $v_i$ are continuous and $\exp$ is continuous, continuity for $W_j$ follows.
\end{proof}

We now show that finding a tuple $\boldl$ where all weight functions are equal gives rise to a $\delta$-robust division.

\begin{lemma}
    \label{lemma:weights-provide-approximate-envy-freeness}
    For a fixed $m \in [0, 1]$, consider a feasible tuple $\boldl$. Suppose that for all $j \in [4]$, $W_{j}(\boldl) = \frac{5}{4}$. Then, $\boldl$ is $\delta$-robust.
\end{lemma}

\begin{proof}
    Consider the $\delta$-favourite graph $G_\delta(\boldl)$.
    By \Cref{lemma:robustness-is-equivalent-to-matching}, $\boldl$ is $\delta$-robust if and only if for every agent $i$, there is a perfect matching in $G_\delta(\boldl)$ minus vertex $i$.
    Now, pick any subset of piece vertices $P$. 
    Then, $\sum_{P_j \in P} W_j - w_{ij} \geq \frac{5}{4} \abs{P} - 1$.
    Every agent for whom a piece $P_j$ is not a $\delta$-favourite, may contribute at most $\exp(\frac{5}{\delta}\cdot (v-\delta))/{\exp(\frac{5}{\delta}\cdot v) = \exp(-\frac{5}{\delta}\delta)} = 2^{-5}$ to $W_j$.

    Hence, the number of agents for whom a piece from $P$ is one of their $\varepsilon$-favourite piece is at least $\ceil{\frac{5}{4}\abs{P} - 1 - 4 \cdot \abs{P} 2^{-5}} \geq \abs{P}$.  So, the condition of \Cref{lemma:robustness-is-equivalent-to-matching} holds.
\end{proof}

We can now use this characterisation of existence along with the continuity and monotonicity properties of $W_{j}(\boldl)$ to prove existence of $\delta$-robust divisions.
For ease of presentation we will extend the valuation functions beyond their domain.

\begin{definition}
    \label{definition:extensions}
    Given a valuation function $v_i \colon [0,1]^2 \rightarrow [0,1]$, we define the valuation's \emph{extension} $\tilde{v_i} \colon \RR^2 \rightarrow \RR_{\geq 0}$ as follows.
     For $a \leq b$, 
    \begin{equation*}
        \tilde{v_i}(a, b) = v_i(\max(0, a), \min(1, b)) + b - a - \max(0, \min(b, 1) - \max(a, 0)).
    \end{equation*}
    That is, $\tilde{v_i}(a, b) = v_i([a, b] \cap [0, 1]) + \mu([a,b] \setminus [0,1])$ where $\mu$ denotes the total length of a union of disjoint intervals.
    For $a > b$, $\tilde{v_i}(a, b) = 0$.
\end{definition}

Importantly, $v_i$ and $\tilde{v_i}$ agree on $[0, 1]^2$: for all $(a, b)\in [0, 1]^2$, $\tilde{v_i}(a, b) = v_i(a, b)$. Further, if $v_i$ is 1-Lipschitz and $\gamma$-strongly hungry, then so is $\tilde{v_i}$.
The benefit of considering extensions $\tilde{v_i}$, is that for every left endpoint $a\in [0,1]$ and target $T\in \RR_{\geq 0}$, there exists $b$ such that $\tilde{v_i}(a, b) = T$.
Given the agreement on the domain $[0,1]^2$, we will often overload $v_i$ to also mean the extension $\tilde{v_i}$.

We are now ready to prove the first existence lemma.
\begin{lemma}
    \label{lemma:existence-of-approximate-robustness}
    For any $m \in [0, 1]$ and $\delta >0$, there exists a maximal feasible tuple $\ellTuple$ that is $\delta$-robust.
\end{lemma}

\begin{proof}
    Fix $m\in[0,1]$ and $\delta>0$, and write $c=5/4$.
    By \Cref{lemma:weights-provide-approximate-envy-freeness}, it suffices
    to find a maximal feasible tuple $\boldl^*$ with $W_j(\boldl^*)=c$
    for every $j\in[4]$.
    Given $\ell_1, \ell_2, \ell_3$ we first find $\ell_4$ such that $W_4(\ell_1, \ell_2, \ell_3, \ell_4) = c$. Then, while keeping the weight of $P_4$ equal to $c$, we continuously move $\ell_3$ until the weight of $P_j$ is also $c$.
    Finally, by keeping the weight of $P_3$ and $P_4$ equal to $c$, we identify a value for $\ell_2$ such that the weight of $P_2$ is also $c$.
    As the weights sum up to 5, the weight of piece $P_1$ is also $c$.
    We show that as we vary $\ell_1$ continuously, the appropriate cuts for all weights to be equal also vary continuously.
    Hence, staring from $\ell_1 = 0$, we can continuously increase $\ell_1$ until either $\ell_1 + \ell_2 = m$ or $\ell_3+\ell_4 = 1-m$. This results in a maximal division that is $\delta$-robust.

    Suppose first that the valuations are hungry. Throughout the
    construction, lengths may be arbitrary nonnegative numbers and
    tuples need not be feasible. We evaluate the pieces using the
    valuation extensions defined in \Cref{definition:extensions}. For each agent,
    the value of each piece is therefore continuous and strictly
    increasing in its length, is zero at length zero, and tends to
    infinity with its length. We use the same formulas for $w_{ij}$
    and $W_j$ on these auxiliary tuples.

    \begin{claim}\label{claim:balance-fourth-piece}
        For every $(\ell_1,\ell_2,\ell_3)\in\mathbb{R}_{\geq0}^3$,
        there is a unique nonnegative length
        $\alpha_4(\ell_1,\ell_2,\ell_3)$ such that
        \[
            W_4\bigl(\ell_1,\ell_2,\ell_3,
                \alpha_4(\ell_1,\ell_2,\ell_3)\bigr)=c.
        \]
        The function $\alpha_4$ is continuous and strictly increasing
        in each argument.
    \end{claim}

    \begin{proof}
        Fix $\vec{u}=(\ell_1,\ell_2,\ell_3)$.
        At $\ell_4=0$, piece $P_4$ has value zero for every agent,
        so $w_{i4}\leq1/4$ and hence $W_4(\vec{u},0)\leq c$.
        As $\ell_4\to\infty$, its value tends to infinity for every
        agent while the other pieces' values stay fixed, so
        $W_4(\vec{u},\ell_4)\to5$.
        Continuity and the intermediate value theorem give a length
        with weight $c$, and strict increase in $\ell_4$ gives uniqueness.
        For a fixed $\ell_4$, increasing any coordinate in $\vec{u}$ strictly decreases $W_4(\vec{u}, \ell_4)$. Hence $\alpha_4(\vec{u})$ is strictly increasing in $\ell_1, \ell_2, \ell_3$.

        To prove continuity, let $r=\alpha_4(\vec{u})$.
        For every $\beta>0$, we have $W_4(\vec{u},r+\beta)>c$.
        This strict inequality persists for inputs $\vec{u}'$
        sufficiently close to $\vec{u}$, giving
        $\alpha_4(\vec{u}')<r+\beta$.
        If $r>0$, we can also choose $0<\beta<r$ and use
        $W_4(\vec{u},r-\beta)<c$ to obtain
        $\alpha_4(\vec{u}')>r-\beta$ for nearby inputs.
        If $r=0$, nonnegativity supplies the lower bound instead.
        These bounds prove continuity, including at a zero solution.
    \end{proof}

    \begin{claim}\label{claim:balanced-family}
        There is a continuous function
        $\vec{x}\colon[0,\infty)\to\mathbb{R}_{\geq0}^4$
        such that $x_1(s)=s$ and $W_j(\vec{x}(s))=c$ for every
        $s\geq0$ and $j\in[4]$.
        Moreover, $\vec{x}(0)=(0,0,0,0)$, and every coordinate
        is strictly increasing and tends to infinity as $s\to\infty$.
    \end{claim}

    \begin{proof}
        We first choose the third length while maintaining $W_4=c$.
        Define
        \[
            F_3(\ell_1,\ell_2,t)
            =W_3\bigl(\ell_1,\ell_2,t,
                \alpha_4(\ell_1,\ell_2,t)\bigr).
        \]
        This function is continuous by \Cref{claim:balance-fourth-piece}.
        Fix $\ell_1,\ell_2$. Increasing $t$ also increases
        $\alpha_4(\ell_1,\ell_2,t)$, so both $W_1$ and $W_2$
        strictly decrease. Since $W_4=c$ and the weights sum to $5$,
        \[
            F_3(\ell_1,\ell_2,t)=3c-W_1-W_2,
        \]
        where the weights on the right are evaluated at the same tuple.
        Thus $F_3$ is strictly increasing in $t$.
        At $t=0$, the third piece has value zero, so $F_3\leq c$.
        As $t\to\infty$, its value tends to infinity for every agent,
        forcing $W_1,W_2\to0$. Hence $F_3\to3c=15/4$.
        The intermediate value theorem and strict increase therefore
        give a unique length $\alpha_3(\ell_1,\ell_2)$ with $F_3(\ell_1, \ell_2, \alpha_3(\ell_1, \ell_2)) = c$.

        At fixed $t$, increasing either $\ell_1$ or $\ell_2$ also
        increases $\alpha_4$. The third piece stays fixed while
        other pieces increase in value, so its weight strictly decreases.
        Thus $F_3$ is strictly decreasing in each of its first two
        arguments, and $\alpha_3$ is strictly increasing in each argument.
        The continuity argument from the preceding claim applies to
        $F_3$ as well, including at a zero root, and proves continuity
        of $\alpha_3$.

        Next, for $s,t\geq0$, define
        \[
            \vec{w}_s(t)=\bigl(
                t,\alpha_3(s,t),
                \alpha_4(s,t,\alpha_3(s,t))\bigr),
            \qquad
            F_2(s,t)=W_2\bigl(s,\vec{w}_s(t)\bigr).
        \]
        Here $W_3=W_4=c$, and consequently
        \[
            F_2(s,t)=2c-W_1\bigl(s,\vec{w}_s(t)\bigr).
        \]
        For fixed $s$, all three coordinates of $\vec{w}_s(t)$
        strictly increase with $t$, so $W_1$ strictly decreases.
        Thus $F_2$ is strictly increasing in $t$.
        At $t=0$, $P_2$ has value zero for all agents, so $F_2(s,0)\leq c$.
        As $t\to\infty$, the value of piece $P_2$ tends to infinity,
        forcing $W_1\to0$ and hence $F_2(s,t)\to2c=10/4$.
        Since $F_2$ is continuous, there is a unique nonnegative
        $\alpha_2(s)$ such that $F_2(s,\alpha_2(s))=c$.

        For fixed $t$, increasing $s$ increases the first, third,
        and fourth lengths, while leaving the second length fixed.
        It therefore strictly decreases $F_2(s,t)$.
        Hence $\alpha_2$ is strictly increasing, and the earlier 
        continuity argument applies to $\alpha_2$.

        We can now set
        \[
            \vec{x}(s)=\bigl(s,\vec{w}_s(\alpha_2(s))\bigr).
        \]
        By construction, $W_2=W_3=W_4=c$, and the constant sum gives
        $W_1=c$ as well. All coordinates of $\vec{x}$ are continuous
        and strictly increasing. At zero input, every root selected
        above is zero, so $\vec{x}(0)=(0,0,0,0)$.

        Finally, $x_1(s)=s\to\infty$.
        If another coordinate $x_j(s)$ remained bounded as $s\to\infty$,
        the value of its piece would remain bounded for each agent,
        while the first piece's value would tend to infinity.
        The weight formula would then give $w_{ij}\to0$ for every agent,
        contradicting $W_j=c$. Since each coordinate is increasing,
        every coordinate must tend to infinity.
    \end{proof}

    We now recover feasibility. By \Cref{claim:balanced-family}, the
    functions $x_1(s)+x_2(s)$ and $x_3(s)+x_4(s)$ are continuous,
    strictly increasing, zero at $s=0$, and unbounded.
    Hence there are unique $s_L,s_R\geq0$ such that
    \[
        x_1(s_L)+x_2(s_L)=m,
        \qquad
        x_3(s_R)+x_4(s_R)=1-m.
    \]
    Set $s^*=\min\{s_L,s_R\}$ and $\boldl^*=\vec{x}(s^*)$.
    Monotonicity gives
    \[
        \ell_1^*+\ell_2^*\leq m,
        \qquad
        \ell_3^*+\ell_4^*\leq1-m,
    \]
    with at least one equality. Thus $\boldl^*$ is maximal feasible,
    and all four weights equal $c$. This also covers $m=0$ and $m=1$,
    for which $s^*=0$.

    Finally, we remove the hungriness assumption. For $\eta\in(0,1)$,
    replace each valuation on intervals $[a,b]\subseteq[0,1]$ by
    \[
        v_i^{\eta}(a,b)=(1-\eta)v_i(a,b)+\eta(b-a).
    \]
    These valuations are hungry, so the construction above gives a
    maximal feasible tuple with all four corresponding weights equal
    to $c$. The set of maximal feasible tuples is closed and bounded.
    We can therefore let $\eta\to0$ along a sequence for which these
    tuples converge to a maximal feasible tuple $\boldl^*$.
    Since $|v_i^\eta(a,b)-v_i(a,b)|\leq\eta$ on every interval in the
    cake, continuity of the original valuations and of the weight
    formula gives $W_j(\boldl^*)=c$ for every $j\in[4]$.
    Applying \Cref{lemma:weights-provide-approximate-envy-freeness}
    completes the proof.
\end{proof}

\begin{lemma}
    \label{lemma:existence-of-exact-robustness}
    Assume that the valuations $v_i$ are hungry. Then, for any $m \in [0, 1]$, there exists a maximal feasible tuple $\ellTuple$ that is robust.
\end{lemma}

\begin{proof}
    To prove this, we can use \Cref{lemma:existence-of-approximate-robustness} and a standard compactness argument similar to that of \cite{Su99-rental-harmony}.
    From \Cref{lemma:existence-of-approximate-robustness}, for every $n \in \ZZ_{+}$ there exists a maximal feasible tuple $\boldl^n$ that is $\frac{1}{n}$-robust.
    By compactness, there is a subsequence $(\boldl^{n_k})$ that converges to a tuple $\boldl^*$.
    As feasibility and maximality are defined by weak inequalities, $\boldl^*$ is a feasible maximal tuple.
    For each tuple $\boldl^{n_k}$, we can construct the $(1/{n_k})$-favourite graph $G_{1/n_k}(\boldl^{n_k})$.
    Notice that there are finitely many possible bipartite graphs with five agent vertices and four piece vertices.
    Hence, there must be a further subsequence $n_{k_r} = p$, such that for each tuple $\boldl^{p}$, $G_{1/p}(\boldl^p)$ is identical for all $p$.
    Denote that graph as $G$.

    As each $\boldl^p$ is $1/p$-robust, the conditions of \Cref{lemma:robustness-is-equivalent-to-matching} hold for $G$.
    Now, consider an edge $(i,P_j)$ of $G$, by continuity:
    \[
        v_i(P_j^*)=\lim_{p\to\infty}v_i(P_j^{p})
        \geq\lim_{p\to\infty}\left(\max_{t\in[4]}v_i(P_t^{p})-\frac{1}{p}\right)
        =\max_{t\in[4]}v_i(P_t^*).
    \]

    Hence, the edge $(i, P_j)$ is present in $G_0(\boldl^*)$ and so $G$ is a subgraph of $G_0(\boldl^*)$.
    Each graph $G_{1/i}(\boldl^i)$ satisfies the matching condition of \Cref{lemma:robustness-is-equivalent-to-matching}. So, $G$ does too. As $G$ is a subgraph of $G_0(\boldl^*)$ so does $G_0(\boldl^*)$.
    Hence, $\boldl^*$ is robust.
\end{proof}

\begin{lemma}\label{lem:uniqueness}
Assume that the valuations are hungry. For any $m \in [0,1]$, there exists at most one maximal feasible tuple $(\ell_1, \ell_2, \ell_3, \ell_4)$ that is robust.
\end{lemma}

\begin{proof}
Towards a contradiction, assume that there exist two distinct such tuples $\boldl$ and $\boldl'$. Without loss of generality, there exists $i \in [4]$ such that $\ell_i > \ell_i'$ (otherwise swap $\boldl$ and $\boldl'$). But then, there must also exist some $j \in [4]$ such that $\ell_j \leq \ell_j'$. Indeed, it is impossible to have $\ell_k > \ell_k'$ for all $k \in [4]$, as then $\boldl$ and $\boldl'$ could not both be maximal feasible.

We have thus argued that the set $S := \{i \in [4]: \ell_i > \ell_i'\}$ satisfies $1 \leq |S| \leq 3$. Now, let $A$ denote the set of agents who have a favourite piece in $S$ in the division described by $\boldl'$. Since $\boldl'$ is robust, we must have $|A| \geq |S| + 1$. Indeed, if there were at most $|S|$ such agents, then removing one of them would make it impossible to assign each piece in $S$ to someone who favours it. Compared to $\boldl'$, in the division corresponding to $\boldl$ we have that the pieces in $S$ have strictly increased in value (since the valuations are hungry), and the pieces in $[4] \setminus S$ have weakly decreased in value, and this is true for all agent valuations. As a result, in the division $\boldl$, each agent in set $A$ now strictly prefers some particular piece in $S$ to any piece in $[4] \setminus S$. This means that the number of agents who have a favourite piece in $[4] \setminus S$ is at most $5 - |A| \leq 5 - (|S| + 1) = 4 - |S|$. This is a contradiction to the robustness of $\boldl$, because if any agent in $[5] \setminus A$ departs, it will not be possible to assign all pieces in $[4] \setminus S$.
\end{proof}

\cref{lemma:existence-of-exact-robustness,lem:uniqueness} imply the following.

\begin{proposition}\label{prop:existence-and-uniqueness}
Assume that the valuations are hungry. For any $m \in [0,1]$, there exists a unique maximal feasible tuple $(\ell_1, \ell_2, \ell_3, \ell_4)$ that is robust.
\end{proposition}

Hence, we can associate with each $m \in [0, 1]$ the unique maximal feasible robust tuple $\boldl^*(m) = (\ell_1^*(m), \ell_2^*(m), \ell_3^*(m), \ell_4^*(m))$.
If $m$ is clear from context, we will occasionally suppress $m$ from $\boldl^*(m)$ and write $\boldl^*$ instead.

\begin{lemma}\label{lem:almost-to-near-robust}
Fix any $m \in [0,1]$ and assume that the valuations are $\gamma$-strongly hungry for some $\gamma \in (0,1)$. If $(\ell_1, \ell_2, \ell_3, \ell_4)$ is a maximal feasible tuple that is $\delta$-robust for some $\delta > 0$, then
\[|\ell_i - \ell_i^*(m)| \leq 3 \gamma^{-3} \delta \qquad \forall i \in \{1,2,3,4\}.\]
\end{lemma}

\begin{proof}
Let $(\ell_1^*, \ell_2^*, \ell_3^*, \ell_4^*)$ denote the unique maximal feasible tuple that is robust at $m$. Let $(\ell_1, \ell_2, \ell_3, \ell_4)$ be a maximal feasible tuple that is $\delta$-robust. Define $d_i := \ell_i - \ell_i^*$ for $i \in \{1,2,3,4\}$. We want to show that $|d_i| \leq 3 \gamma^{-3} \delta$. Let $I^+$ denote the set of indices $i \in \{1,2,3,4\}$ such that $d_i > 0$. Note that we must have $|I^+| \leq 3$. Indeed, if we had $\ell_i > \ell_i^*$ for all $i \in \{1,2,3,4\}$, then the tuple $\bm{\ell}$ could not be maximal feasible, as the inequality that holds with equality for $\bm{\ell^*}$ would not hold at all for $\bm{\ell}$.

When we go from $\bm{\ell^*}$ to $\bm{\ell}$, for each piece $i \in I^+$, its value to any agent increases by at most $d_i$, and by at least $\gamma d_i$. This is due to the $1$-Lipschitzness of the valuations and to the fact that they are $\gamma$-strongly-hungry. The value of any other piece, i.e., any piece $i \in \{1,2,3,4\} \setminus I^+$, possibly decreases, but certainly does not increase for any agent.

Now denote $d_{\max} := \max_{i \in I^+} d_i$ and consider the line $[0, d_{\max}]$. The intervals $[\gamma d_i, d_i]$, $i \in I^+$, cover some parts of that line. Let us now argue that the union of these intervals cannot leave a subinterval of $[0, d_{\max}]$ of length strictly larger than $\delta$ uncovered. Towards a contradiction, assume that there is such an interval $(a,b)$ with $b-a > \delta$ and $(a,b) \subseteq [0, d_{\max}] \setminus \cup_{i \in S} [\gamma d_i, d_i]$. Let $S$ denote the set of indices $i \in I^+$ such that the interval $[\gamma d_i, d_i]$ lies above the uncovered interval $(a,b)$, i.e., $\gamma d_i \geq b$. Note that $S \neq \emptyset$, because there exists $i \in I^+$ with $d_i = d_{\max}$. Furthermore, we also have that $|S| \leq 3$, since $S \subseteq I^+$.

Since the tuple $\bm{\ell^*}$ is robust, it follows that, when the cuts are positioned according to $\bm{\ell^*}$, there exist at least $|S|+1$ agents who weakly prefer a piece in $\{P_i\}_{i \in S}$ to any piece in $\{P_i\}_{i \notin S}$. Indeed, if there were at most $|S|$ such agents, then after the departure of one of them, it would not be possible to assign each piece in $\{P_i\}_{i \in S}$ to some agent who likes it, a contradiction to robustness. Having argued that at least $|S|+1$ such agents exist, let us denote by $A$ this set of agents. Now, if we instead position the cuts according to $\bm{\ell}$, the agents in $A$ prefer some piece in $\{P_i\}_{i \in S}$ by more than $\delta$ compared to any piece in $\{P_i\}_{i \notin S}$, i.e.,
\begin{equation}\label{eq:almost-to-near-robust:delta-gap}
\max_{i \in S} v_j(P_i) > \max_{i \notin S} v_j(P_i) + \delta \qquad \forall j \in A.
\end{equation}
Indeed, when we go from $\bm{\ell^*}$ to $\bm{\ell}$, for any agent we have that the value of any piece in $\{P_i\}_{i \in S}$ increases by at least $b$, and the value of any piece in $\{P_i\}_{i \notin S}$ increases by at most $a$. Since $b-a > \delta$, \eqref{eq:almost-to-near-robust:delta-gap} follows. However, \eqref{eq:almost-to-near-robust:delta-gap} contradicts the $\delta$-robustness of $\bm{\ell}$, as it says that no piece in $\{P_i\}_{i \notin S}$ can be assigned to any agent in $A$ with envy at most $\delta$. This means that there are at most $5 - |A| \leq 5 - (|S|+1) = 4-|S|$ agents who can receive one of the $4 - |S|$ pieces in $\{P_i\}_{i \notin S}$. As soon as one of these agents departs, some piece in $\{P_i\}_{i \notin S}$ will have to remain unassigned, a contradiction.

We have thus shown that $\cup_{i \in S} [\gamma d_i, d_i]$ leaves only intervals of length at most $\delta$ uncovered in $[0, d_{\max}]$. From this, it is not hard to deduce an upper bound on the value of $d_{\max}$. Indeed, since $|S| \leq 3$, the worst-case is achieved by having the three intervals $[\delta, \gamma^{-1}\delta]$, $[\gamma^{-1}\delta + \delta, \gamma^{-1}(\gamma^{-1}\delta + \delta)]$, and $[\gamma^{-1}(\gamma^{-1}\delta + \delta) + \delta, \gamma^{-1}(\gamma^{-1}(\gamma^{-1}\delta + \delta) + \delta)]$. This gives the bound $d_{\max} \leq \gamma^{-1}(\gamma^{-1}(\gamma^{-1}\delta + \delta) + \delta) \leq 3 \gamma^{-3} \delta$.

This shows that $d_i \leq 3 \gamma^{-3} \delta$ for all $i \in \{1,2,3,4\}$. To show that $d_i \geq - 3 \gamma^{-3} \delta$, consider the set $I^-$ of indices $i$ such that $d_i < 0$, and use a very similar argument.
\end{proof}

\begin{lemma}\label{lem:lipschitzness-wrt-m}
Assume that the valuations are $\gamma$-strongly hungry for some $\gamma \in (0,1)$. Then, the function $m \mapsto (\ell_1^*(m), \ell_2^*(m), \ell_3^*(m), \ell_4^*(m))$ is $(13 \gamma^{-3})$-Lipschitz-continuous.
\end{lemma}

\begin{proof}
Consider $m, m' \in [0,1]$ with $|m-m'| =: d > 0$. Our goal is to show that $|\ell_i^*(m) - \ell_i^*(m')| \leq 13 \gamma^{-3} d$ for all $i \in [4]$. First, we argue that there exists a tuple $\boldl$ that has the following two properties:
\begin{itemize}
\item it is close to $\boldl^*(m)$, namely $|\ell_i - \ell_i^*(m)| \leq d$ for all $i \in [4]$, and
\item it is maximal feasible at $m'$.
\end{itemize}
It is not hard to see that such a tuple must indeed exist. Namely, it suffices to initially set $\ell_i := \ell_i^*(m)$ for all $i \in [4]$, and then to slightly move the $\ell_i$ to recover maximal feasibility at $m'$. Since $|m-m'| = d$ and $\boldl^*$ is maximal feasible at $m$, it suffices to move each $\ell_i$ by at most $d$.

Now, since $|\ell_i - \ell_i^*(m)| \leq d$ for all $i \in [4]$, and $|m-m'| = d$, it follows that the value of any piece for any agent can only differ by at most $2d$ in the division defined by $(m', \boldl)$ compared to the division defined by $(m, \boldl^*(m))$. Here, we used the fact that all valuations are $1$-Lipschitz-continuous. Since $\boldl^*(m)$ is robust at $m$, it thus follows that $\boldl$ is $4d$-robust at $m'$.

Finally, by \cref{lem:almost-to-near-robust} we obtain that $|\ell_i - \ell_i^*(m')| \leq 3 \gamma^{-3} \cdot 4d = 12 \gamma^{-3} d$. As a result, it holds that for all $i \in [4]$
\[|\ell_i^*(m) - \ell_i^*(m')| \leq |\ell_i^*(m) - \ell_i| + |\ell_i - \ell_i^*(m')| \leq d + 12 \gamma^{-3} d \leq 13 \gamma^{-3} d. \qedhere\]
\end{proof}

Given an anchor $m$ and a feasible tuple $\boldl$, we can define 6 pieces.
We have already discussed pieces $P_1, \ldots, P_4$. Additionally, we let $L = [\ell_1, m - \ell_2]$ and $R = [m + \ell_3, 1 - \ell_4]$.
Notice that if $\boldl$ is maximal, one of $L$ and $R$ is empty.

\begin{lemma}
    \label{lemma:robustness-plus-gives-envy-freeness}
    Suppose that for an anchor $m \in [0, 1]$ a tuple $\boldl$ is feasible, maximal and robust.
    Suppose that for some agent $i$, piece $L$ or piece $R$ is one of agent's favourite pieces.
    Suppose that for all other agents $i' \neq i$, one of pieces $P_1, \ldots, P_4$ is the agent $i'$'s favourite.
    Then, there is an exact envy-free complete allocation for five agents.
\end{lemma}

\begin{proof}
    By maximality of $\boldl$, either $L$ or $R$ is empty.
    Without loss of generality, let $L$ be empty.
    Then, we can assign piece $R$ to agent $i$. Piece $R$ is one of $i$'s favourite pieces, so he does not envy any other agents.
    For every other agent $i' \neq i$, one of the $P_j$ are agent $i'$'s favourite pieces, so no agent envies $i$.
    By the robustness of $\boldl$, there is a way to allocate pieces $P_1, \ldots P_4$ to the agents that are not $i$, such that no agent envies the piece of a different agent.
    Hence, there is an envy-free allocation.
\end{proof}

We are now ready to prove the existence of an envy-free division for five agents.
\theoremExistence*

\begin{proof}
    First, suppose that the valuations $v_i$ are $\gamma$-strongly hungry.
    Let $\boldl^*(m)$ the maximal feasible tuple that is robust for the anchor $m \in [0,1]$.
    Then, by \Cref{lem:lipschitzness-wrt-m}, $\boldl^*(m)$ is continuous in $m$.
    Denote as $L$ the piece $[\ell_1, m - \ell_2]$ and by $R$ the piece $[m + \ell_3, 1 - \ell_4]$.
    Notice that for a maximal $\boldl$, one of $L$ and $R$ must be empty.
    For $m = 0$, $\boldl^*(m)=(0,0,0,0)$.
    In this case, piece $R$ is strictly preferred over the empty pieces $P_1, \ldots, P_4, L$.
    For $m = 1$, it is also the case that $\boldl^*(m)=(0,0,0,0)$.
    In this case, piece $L$ is strictly preferred over the other pieces $P_1, \ldots, P_4, R$.
    By the intermediate value theorem, there must be a value $m^*$ such that for all agents $i$, piece $R$ is not the strict favourite and for some agent $i$ it is a jointly favourite piece.
    Hence, we can apply \Cref{lemma:robustness-plus-gives-envy-freeness} to deduce that there is an envy-free allocation of pieces $P_1, \ldots, P_4, R$ to agents $1, 2, \ldots, 5$.

    Now, we dispense of the $\gamma$-strong hungriness assumption using a standard compactness argument.
    Consider valuations $v_i$ that need not be $\gamma$-strongly hungry.
    For each integer $n \geq 2$, let $v_i^n(a, b) = (1 - 1/n)v_i(a, b) + (1/n)\max(b -a ,0)$.
    Then, for each $v_i^n$ we can compute an exact envy-free allocation given by $(m_n, \boldl^*(m_n))$.
    Each allocation maps a piece from $P_1, \ldots, P_4, R$ to agents $1, \ldots, 5$.
    As the number of such mappings is finite, one such allocation must appear infinitely many times.
    Consider the subsequence $m_k$ where only that allocation is appears.
    As the possible values of $(m_k, \boldl^*(m_k))$ form a compact space, there is a subsequence $(m_r, \boldl^*(m_r))$ that converges to some $(m^*, \boldl^*)$ as $r \to \infty$.
    
    Let $P^*$ the interval of piece $P$ in the division $(m^*, \boldl^*)$.
    For each piece $P$, let $P^r$ the interval it is assigned in division $(m_r, \boldl^*(m_r))$.
    Then, $\abs{v_i(P^r) - v_i^r(P^r)} \leq 1/r$.
    Also, given two intervals $P = [a, b], Q = [c, d]$ let $\abs{P - Q} = \abs{a - c} + \abs{b - d}$.
    Then, as $r \rightarrow \infty$, for all pieces $P$, $\abs{P^* - P^r} \rightarrow 0$.
    As the $v_i$ are continuous, $v_i(P^r) \to v_i(P^*)$.
    Hence, as $r \to \infty$, $v_i^r(P^r) \rightarrow v_i(P^*)$.
    As limits preserve weak inequalities and envy-freeness is defined using weak inequalities, the division $(m^*, \boldl^*)$ is envy-free.
\end{proof}

\section{An Efficient Algorithm for Five Agents}

We can use the existence proof and the machinery we developed to extract an efficient algorithm that uses $\poly(\log(1/\varepsilon))$ queries.

\begin{restatable}{theorem}{theoremEfficientAlgorithm}
    \label{theorem:efficient-algorithm}
    Consider an instance of cake-cutting with Lipschitz continuous and monotone valuation functions $v_i$. Then, for every $\varepsilon>0$, we can compute an $\varepsilon$-envy-free complete allocation for five agents with contiguous pieces using $\mathcal{O}(\log^3(\varepsilon^{-1}))$ valuation queries.
\end{restatable}

The backbone of our algorithm will be repeated use of interval bisection.
The following standard lemma will prove useful.

\begin{lemma}
    \label{lemma:interval-bisection}
    Assume that the valuations $v_i$ are hungry.
    For a given $m \in [0, 1]$, suppose there exists $\ell_j^* \in (0, 1)$ such that for some agent $i$ and some value $x$, $v_i(P_j^*) = x$.
    Then, given $x$, for any precision $p > 0$, there is an algorithm for finding a length $\ell_j$ such that $\abs{\ell_j - \ell_j^*} \leq p$, in time $\mathcal{O}(\log(1/p))$. 
    In particular, we can also require that $\ell_j \leq \ell_j^*$. For such an $\ell_j$, we also have that $\abs{v_i(P_j) - v_i(P_j^*)} \leq p$.
    
\end{lemma}

\begin{proof}
    This can be proven directly using iterated interval bisection.
    For ease of notation we will consider the case where $j = 1$.
    As $v_i$ is a strictly increasing normalised function, we know that $v_i(0, 0) = 0 < x <v_i(0,1)$.
    Given any interval $[a, b]$ where $v_i(0, a) \leq x \leq v_i(0, b)$, we can find the midpoint $\mathit{mid} = (a + b)/2$. If $v_i([0, \mathit{mid}]) \leq x$, retain $[\mathit{mid},b]$. Otherwise, retain $[a,\mathit{mid}]$.
    In either case, the length of the interval is halved, but the property that $v_i(0, a) \leq x \leq v_i(0, b)$ is maintained.
    Initially $a = 0$ and $b = 1$.
    For any positive precision $p > 0$, after $\mathcal{O}(\log(1/p))$ iterations, the length of the interval is at most $p$. 
    As $\ell_j^* \in [a, b]$, $a \leq \ell_j^* \leq a + p$. 
    Hence, we can select $a$ as $\ell_j$.
    Also, by the fact that $v_i$ are $1$-Lipschitz, $\abs{v_i(0, \ell_j) - v_i(0, \ell_j^*)} \leq \abs{\ell_j - \ell_j^*} \leq p$. 
\end{proof}

For a fixed $m \in [0,1]$, there exists a unique maximal feasible robust tuple $\boldl^*(m)$. Given this tuple, we can easily compute the exact favourite graph $G_0(\boldl^*(m))$.
We can recover a partial converse: given a spanning tree $T$ of $G_0(\boldl^*(m))$, for every $\delta > 0$, we can recover a tuple $\boldl$ such that $T$ is a subgraph of $G_{3\delta}(\boldl)$.
This process requires polylogarithmic queries on $1/\delta$ and $1/\gamma$.

\begin{lemma}
    \label{lemma:tree-to-approximately-robust}
    Assume that the valuations are $\gamma$-strongly hungry for some $\gamma > 0$. Consider a connected bipartite graph $G$ with vertex set $[5] \sqcup \{P_1, P_2, P_3, P_4\}$ and a fixed $m \in [0, 1]$.
    Let $T$ be a spanning tree of $G$.
    Suppose there exists a maximal feasible robust tuple $\boldl^*$ such that  $T$ is a subgraph of the exact favourite graph $G_0(\boldl^*)$.
    Then, for any $\delta >0$, we can find a maximal feasible $3\delta$-robust $\boldl$ such that $T$ is a subgraph of $G_{3\delta}(\boldl)$ with $\mathcal{O}(\log(\delta^{-1}\gamma^{-1}) \cdot \log(\delta^{-1}))$ queries.
    Additionally, we can ensure that $\boldl$ is close to $\boldl^*$:
    \[|\ell_j - \ell_j^*| \leq \delta \qquad \forall j \in \{1,2,3,4\}.\]
\end{lemma}
\begin{proof}
    Let $T$ be a spanning tree of a bipartite graph $G$ with vertex set $[5] \sqcup \{P_1, \ldots, P_4\}$.
    Consider an agent vertex $i$ with degree at least 2 in $T$.
    Then, for $T$ to be a subgraph of $G_0(\boldl^*)$ we must have that for all pieces $P_j^*, P_{j'}^*$ adjacent to $i$, $v_i(P_j^*) = v_i(P_{j'}^*)$.
    
    We perform iterated interval bisection on the interval $[0, m]$ to find an appropriate value of $\ell_1$.
    For a candidate $\ell_1$, we perform the following.
    Suppose that for agent $i$, pieces $P_1$ and $P_j$ are adjacent in $T$.
    By considering the extension of $v_i$ as defined in \Cref{definition:extensions}, there always exists $\ell_j'$ such that $v_i(P_j') = v_i(P_1)$.
    By \Cref{lemma:interval-bisection}, for all $p > 0$, we can find $\ell_j$ such that $\ell_j \leq \ell_j'$ and $\abs{\ell_j -\ell_j'} \leq p$ in $\mathcal{O}(\log(1/p))$ time.
    Once we fix a candidate $\ell_j$, we can perform the same operation to any other piece $P_{j'}$ for which there exists an agent $i'$ for which pieces $P_1$ or $P_j$ are adjacent to $P_{j'}$.
    As $T$ is a spanning tree, every piece is reachable by this process, so we construct a candidate tuple $\boldl$.
    This can be done in $\mathcal{O}(\log(1/p))$ queries as the interval bisections are \emph{sequential} and not nested.

    As such, we have constructed a candidate tuple $\boldl$.
    Suppose that $\ell_1 \leq \ell_1^*$. Then, by monotonicity of the $\ell_j$, we must have that $\ell_j \leq \ell_j^*$.
    In particular, if the tuple $\boldl$ is infeasible, $\ell_1$ overshoots $\ell_1^*$ and we have that $\ell_1 > \ell_1^*$.
    Suppose instead that for some $K > 0$, $\ell_1 \geq \ell_1^* + K$.
    Then, by $\gamma$-strong hungriness, $v_i(P_1) \geq v_i(P_1^*) + K\gamma$.
    Hence, $v_i(P_j) + p \geq v_i(P_1) \geq v_i(P_1^*) + K \gamma$.
    As for $\ell_j^*$ $v_i(P_j^*) = v_i(P_1^*)$, we deduce $v_i(P_j) + p \geq  v_i(P_j^*) + K \gamma$. 
    If $K\gamma -p \geq 0$, from $1$-Lipschitz continuity this implies that $\ell_j \geq \ell_j^* + K\gamma - p$.
    This argument can be generalised: suppose $\ell_j \geq \ell_j^* + K$ and for some agent $i$, pieces $j, j'$ are neighbours of $i$ in $T$.
    Then, $\ell_{j'} \geq \ell_{j'}^* + K\gamma - p$.
    As the maximum distance in the tree is $3$, we can conclude if $K\gamma -p\gamma^2-p\gamma - p >0$, then for all pieces $P_j$,  $\ell_j \geq \ell_j^* + K\gamma^3 - p \gamma^2 - p \gamma - p$. As $\gamma \leq 1$, we can rewrite this as: $\ell_j \geq \ell_j^* + K \gamma^3 - 3p$.
    In particular, for $K > 3 p \gamma^{-3}$, $\boldl$ will be infeasible as $\boldl^*$ is maximal.
    This allows us to perform the following test.
    If $\boldl$ is infeasible then $\ell_1 > \ell_1^*$.
    If $\boldl$ is feasible then $\ell_1 \leq \ell_1^*+ 3p\gamma^{-3}$.

    Using this test, we can perform interval bisection.
    Set $p = \delta\gamma^3/6$.
    We maintain an interval of possible $\ell_1$ that we denote as $[a_1, b_1]$.
    Initially, $[a_1, b_1] = [0, m]$.
    We select as candidate $\ell_1$ the midpoint $(a_1 + b_1)/2$.
    We construct a candidate tuple $\boldl$ as described earlier in this proof.
    If $\boldl$ is feasible, then we set $a_1 = \ell_1$. If it is infeasible, we set $b_1 = \ell_1$.
    This ensures that at each step of the algorithm, $b_1 \geq \ell_1^*$ and $a_1 \leq \ell_1^*+ \delta/2$.
    After $\log(1/p)$ iterations, the length of the interval $[a_1, b_1]$ is $p = \delta \gamma^3/6 \leq \delta/2$.
    Then, $b_1 - a_1 \leq \delta/2$ and so $a_1 - \delta/2 \leq \ell_1^* \leq a_1 + \delta/2$.
    This argument is not unique to piece $P_1$ and we can apply it to pieces $P_2, P_3, P_4$ to find $a_2, a_3, a_4$ such that $a_j - \delta/2 \leq \ell_j^* \leq a_j + \delta/2$.
    Select $\ell_j = a_j - \delta/2$. Then, $\boldl$ is a feasible allocation with $\norm{\boldl - \boldl^*}_\infty \leq \delta$. 
    As the $v_i$ are $1$-Lipschitz and $\boldl^*$ is robust, $\boldl$ is $\delta$-robust.

    The final step is to increase one $\ell_j$ to recover maximality.
    As $\boldl^*$ is maximal, either $\ell_1^* + \ell_2^* = m$ or $\ell_3^* + \ell_4^* = 1-m$.
    In particular, either $\ell_1 + \ell_2 \geq m - 2 \delta$ or $\ell_3 + \ell_4 \geq 1-m -2 \delta$.
    If it is the first case, we increase $\ell_2$ to $m-\ell_1$. In the second case, we increase $\ell_3$ to $1-m-\ell_4$.
    In either case the length is increased by at most $2\delta$.
    By 1-Lipschitzness, the value of $P_2$ or $P_3$ respectively, increases by at most $2\delta$.
    Hence, the division is $3\delta$-robust.
    In particular, $T$ is a subtree of $G_{3\delta}(\boldl)$.

    As for the running time, we require $\mathcal{O}(\log(\delta^{-1}\gamma^{-1}))$ queries for the external interval bisection and $\mathcal{O}(\log(\delta^{-1}\gamma^{-1}))$ queries for the internal bisection, resulting in $\mathcal{O}(\log^2(\delta^{-1}\gamma^{-1}))$ queries.
\end{proof}

As the number of bipartite graphs with five agents and four pieces is constant, and for each graph the number of spanning trees is constant, the possible number of spanning trees of such graphs is a constant.
As such, we can ``guess'' the correct spanning tree by iterating through the constant number of spanning trees.

\begin{lemma}\label{lem:approx-robust-cuts}
Assume that the valuations are $\gamma$-strongly hungry for some $\gamma > 0$. Given any anchor $m \in [0,1]$, we can compute a maximal feasible $\delta$-robust tuple $(\ell_1, \ell_2, \ell_3, \ell_4)$ that satisfies
\[|\ell_j - \ell_j^*(m)| \leq \delta \qquad \forall j \in \{1,2,3,4\}\]
using $\mathcal{O}(\log^2(\delta^{-1}\gamma^{-1}))$ value queries to the valuation functions.
\end{lemma}

\begin{proof}
    We can apply \Cref{lemma:tree-to-approximately-robust} directly.
    For each $m \in [0,1]$, there exists a maximal robust feasible division $\boldl^*$.
    This division defines a favourite graph $G_0(\boldl^*)$ with a spanning tree $T$.
    By \Cref{lemma:tree-to-approximately-robust}, given $T$, for any $p >0$ we can compute $\boldl$ such that $\boldl$ is a $3p$-robust division in $\mathcal{O}(\log(p^{-1}) \log(p^{-1}\gamma^{-1}))$ time.
    The number of trees $T$ in a bipartite graph with 5 agent vertices and 4 piece vertices is constant.
    So, we run the algorithm on the constant number of spanning trees and efficiently verify if an allocation is $3p$-robust.
    Hence, we are guaranteed to find a $3p$-robust division.
    By \Cref{lem:almost-to-near-robust}, that means that for all $j \in [4]$, $\abs{\ell_j - \ell_j^*(m)} \leq 9p\gamma^{-3}$.
    By setting $p = \gamma^3 \delta/9$, we can find $\ell_j$ such that $\abs{\ell_j - \ell_j^*(m)} \leq \delta$ in time $\mathcal{O}(\log^2(\delta^{-1}\gamma^{-1}))$.
\end{proof}

Given an anchor $m \in [0,1]$ we can efficiently compute a maximal $\delta$-robust tuple $\boldl$.
We can use this as a subroutine to find an $\varepsilon$-envy-free division.
For a given tuple $\boldl$, let $L$ denote the piece $[\ell_1, m - \ell_2]$ and $R$ be the piece $[m + \ell_3, 1 - \ell_4]$.
Notice that at a maximal division, at least one of $L$ and $R$ is empty.
As we argued in the proof of \Cref{theorem:existence-of-ef-5},
there must exist $m^*$ such that $\boldl^*(m^*)$ is an exact envy-free allocation.
Using Lipschitz continuity and $\gamma$-strong hungriness, we can replicate this argument to find an approximate envy-free division.

\begin{lemma}
    \label{lemma:efficient-algorithm-for-hungry-agents}
    Assume that the valuations are $\gamma$-strongly hungry for some $\gamma > 0$.
    For every $\varepsilon>0$, we can compute an $\varepsilon$-envy-free complete division for five agents with contiguous pieces using $\mathcal{O}(\log^3(\varepsilon^{-1}\gamma^{-1}))$ valuation queries.
\end{lemma}
\begin{proof}
    We start by performing interval bisection to compute a suitable $m$.
    We maintain an interval of possible $m$, which we denote as $[a, b]$.
    The initial interval is $[0,1]$.
    At each iteration we select as a candidate $m$ the midpoint $m = (a + b)/2$.
    For a candidate $m$, we compute $\boldl$ such that $\norm{\boldl -\boldl^*(m)}_\infty \leq \varepsilon /12$.
    By \Cref{lem:approx-robust-cuts}, this can be done in $\mathcal{O}(\log^2(\varepsilon^{-1}\gamma^{-1}))$ queries.
    If there is an agent $i$ for which piece $R$ is one of his favourite pieces, set $a = m$. Otherwise, set $b = m$.

    We can ensure that $\abs{a - b} \leq \varepsilon\gamma^3/(13\cdot 12)$ after $ \mathcal{O}(\log(\varepsilon^{-1}\gamma^{-1}))$ queries.
    But, $\boldl^*(m)$ is $13\gamma^{-3}$-Lipschitz continuous by \Cref{lem:lipschitzness-wrt-m}.
    As such, $\norm{\boldl^*(a) - \boldl^*(b)}_\infty \leq \varepsilon/12$.
    Let $\boldl_a$ be the tuple computed for $a$ and $\boldl_b$ be the tuple computed for $b$.
    By our earlier construction, $\norm{\boldl_a -\boldl^*(a)}_\infty \leq \varepsilon/12$ and $\norm{\boldl_b -\boldl^*(b)}_\infty \leq \varepsilon/12$. By the triangle inequality, $\norm{\boldl_a - \boldl_b}_\infty \leq \varepsilon /4$.
    Let $P_a$ be one of the five non-empty pieces in $\boldl_a$ and $P_b$ be the corresponding piece in $\boldl_b$.
    By $1$-Lipschitzness of the valuation functions, for all agents $i$ and pieces $P_a$ with corresponding piece $P_b$, $\abs{v_i(P_a) - v_i(P_b)} \leq \varepsilon/2$.
    In particular, if a piece $P_a$ is a favourite piece for agent $i$ in division $\boldl_a$, then $P_b$ must be $\varepsilon$-favourite piece in division $\boldl_b$.
    As $R$ is a favourite piece for some agent $i$ in $\boldl_a$, it is an $\varepsilon$-favourite piece for agent $i$ in $\boldl_b$.

    We now claim that $\boldl_b$ is an $\varepsilon$-envy-free division.
    Suppose the interval $R_b$ has positive length.
    Then, we can first assign piece $R$ to agent $i$ for whom it is an $\varepsilon$-favourite.
    For no agent is $R$ a favourite piece in $\boldl_b$, so no agent envies $i$.
    By the fact that $\boldl_b$ is $\varepsilon/12$ robust, there is an allocation of pieces $P_1, \ldots, P_4$ to agents $[5] \setminus \{i\}$ such that every agent receives an $\varepsilon/12$-favourite piece among $P_1, \ldots, P_4$.
    As $R$ is not a favourite piece, this is an $\varepsilon$-envy-free division.
    Now, consider the edge case where $R_b$ has length $0$ in $\boldl_b$. If piece $L_b$ is non-empty, we can assign piece $L_b$ to agent $i$ and so completely allocate the cake.
    Piece $L_b$ has strictly positive length, so it is also an $\varepsilon$-favourite piece of agent $i$.
    If piece $L_b$ is also empty, then we can allocate either empty piece to agent $i$ while maintaining $\varepsilon$-envy-freeness.

    The total number of queries (and runtime of the algorithm) is $\mathcal{O}(\log^3(\varepsilon^{-1}\gamma^{-1}))$.
\end{proof}

With a very short argument, we can dispense the assumption of $\gamma$-strong hungriness.
Doing so proves \Cref{theorem:efficient-algorithm} and concludes the algorithmic section.

\theoremEfficientAlgorithm*

\begin{proof}
    As noted in the preliminaries, given a valuation $v_i$ and for any $\gamma \in (0, 1)$, we can construct a $\gamma$-strongly hungry valuation $v_i'$ by setting $v_i'(\ell, r) = (1-\gamma)v_i(\ell, r) + \gamma\max(r- \ell, 0)$.
    As for all $(\ell, r) \in [0, 1]^2$, $\abs{v_i(\ell, r) - v_i'(\ell, r)} \leq \gamma$, any $\delta$-envy-free allocation with respect to $v_i'$ is a $(\delta + 2\gamma)$-envy-free allocation with respect to the original $v_i$.

    Hence, we can directly apply \Cref{lemma:efficient-algorithm-for-hungry-agents}.
    Given valuations $v_i$ and target input $\varepsilon$, construct $\varepsilon/3$-strongly hungry valuations $v_i'$ and search for an $\varepsilon/3$-envy-free allocation.
    By \Cref{lemma:efficient-algorithm-for-hungry-agents}, this can be done in $\mathcal{O}(\log^3(\varepsilon^{-2})) = \mathcal{O}(\log^3(1/\varepsilon))$ valuation queries.
\end{proof}

\newpage
\bibliography{bib.bib}
\bibliographystyle{plainnat}
\end{document}